\documentstyle[12pt,epsf]{article}
\newcommand{\ds }{\displaystyle}
\newcommand{\ra}{\rightarrow}
\newcommand{\be}{\begin{equation}}
\newcommand{\ee}{\end{equation}}
\newcommand{\bea}{\begin{eqnarray}}
\newcommand{\eea}{\end{eqnarray}}
\newcommand{\ci}{\cite}
\newcommand{\bi}{\bibitem}
\newcommand{\nono}{\nonumber \\}

\newcommand{\quart}{\frac{1}{4}}

\newcommand{\dd}{\partial}

\newcommand{\eps}{\epsilon}

\newcommand{{\bfna}}{\mbox{\boldmath$\vec{\nabla}$}}

\newcommand{\half}{\frac{1}{2}}

\def\dal{\,\lower0.3ex\vbox{\hrule\hbox{\vrule\kern2pt\vbox{\kern4pt\kern4pt}
\kern2pt\vrule}\hrule}\,}

\begin{document}

\title{\sl Decay of a square pulse to Sine-Gordon breathers}
\vspace{1 true cm}
\author{G. K\"albermann$^*$
\\Soil and Water dept., Faculty of
Agriculture, Rehovot 76100, Israel}
\maketitle

\vspace{3 true cm}
\begin{abstract}

We investigate numerically and analytically 
the existence of thresholds for the production of 
Sine-Gordon single and multiple breathers from simple initial pulses. 
\end{abstract}

$^*${\sl e-mail address: hope@vms.huji.ac.il}

\newpage
\section{\sl Introduction}

The Sine-Gordon equation arises in physical problems such as
one dimensional dislocations
\ci{lamb}, long Josephson junctions\ci{rem}, as well as in the mathematical
treatment of constant negative curvature metric spaces\ci{eisenhart}.

Solitary wave solutions of the Sine-Gordon equation carry a 
topological winding number {\sl q}. 
The solitary waves conserve their properties upon collision and are therefore 
called topological solitons. 
The winding number zero sector {\sl q=0} consists of bound soliton-antisoliton
solutions, the breathers, unbound soliton-antisoliton 
pairs\ci{lamb}, and phonons. The latter are small amplitude solutions, 
corresponding to the Klein-Gordon linear limit of the equation.
The {\sl q=1} sector solitary wave is the shelf or kink soliton.

The Sine-Gordon equation is one of a gallery of nonlinear
equations amenable to treatment by the Inverse Scattering Transform (IST)
 method. The IST produces soliton solutions based on scattering data of 
Dirac-like equations for a potential connected to the sought
solution. The wave functions and the scattering data determine
the potential - hence the name
inverse scattering transform - through the integral equations named after
Gelfand-Levitan and Marchenko\ci{ablowitz}.

The power of the IST method lies in the mapping of a nonlinear
problem to an eigenvalue Sturm-Liouville problem.
We recently used the IST method to treat the decay of 
distorted kinks to wobble states and phonons. \ci{k1}
In the present work we investigate the decay of a square pulse
to single and multiple breathers and phonons. 
The IST technique determines the constraints on the parameters 
of an initial pulse for it to decay into breathers and phonons.
These constraints are satisfied at the location of a 
pole in the transmission amplitude of the IST scattering problem.

The decay of simple pulses to breathers was addressed 
some time ago by Kivshar and coworkers \ci{kiv1,kiv2}.
Kivshar et al.\ci{kiv1,kiv2} studied the creation of breathers 
in long Josephson junctions and easy plane ferromagnets with
an inhomogeneous Sine-Gordon equation.
With a specific choice of the inhomogeneous source term 
and under the approximation of a negligible initial field configuration,
the Zakharov-Shabat (ZS) equations\ci{zakh,akns} are simplified considerably 
and an analytical solution of Kaup\ci{kaup} could be implemented.
In the present work we will focus on the homogeneous
Sine-Gordon equation with a non negligible initial pulse.

In section 2 we treat the decay of square pulse analytically using the
AKNS\ci{akns} version of the Zakharov-Shabat\ci{zakh} equations. The results
are compared to a full numerical treatment of the equations with a more
realistic pulse.
Section 3 provides numerical support to the analytical results of section 2.
Conclusions are drawn in section 4.

\section{\sl Decay of a square pulse in the IST}

The Sine-Gordon equation is usually expressed in two ways. 
In space-time coordinates 

\be\label{sg}
\frac{\dd^2 u}{\dd T^2}-\frac{\dd^2 u}{\dd X^2} + sin(u) = 0.
\ee

\noindent and in light-cone coordinates $\ds x=\half (X+T)~t=\half (X-T)$
with unit speed of light $\ds c=1$,

\be\label{sg1}
\frac{\dd}{\dd t}~\bigg(\frac{\dd u}{\dd x}\bigg) - sin(u) = 0.
\ee
\noindent The Klein-Gordon equation for a particle of unit mass
is obtained by linearizing Eqs.(\ref{sg},\ref{sg1}).
Eq.(\ref{sg}) can be derived from the lagrangian

\bea\label{lag}
{\cal L}=\int{~dX~\bigg(\bigg(\frac{\dd u}{\dd T}\bigg)^2
-\bigg(\frac{\dd u}{\dd X}\bigg)^2+(cos(u)-1)}\bigg),
\eea
\noindent

\noindent possesing a conserved energy
\bea\label{en}
{\cal E}=\int{~dX~\bigg[\bigg(\frac{\dd u}{\dd T}\bigg)^2
+\bigg(\frac{\dd u}{\dd X}\bigg)^2-(cos(u)-1)}\bigg].
\eea

The IST method uses solutions of the Sine-Gordon and
other nonlinear equations as a potential in 
a linear scattering problem of an anciliary wave function. 
The scattering problem is split into a time independent
set of equations and a time evolution set.
For the space-time formulation of the Sine-Gordon equation the equations are
\ci{akns}

\bea\label{akns}
v_{1,X}&=&-i(\frac{\xi}{2}-\frac{cos(u)}{8~\xi}) v_1
+(\frac{i}{8~\xi}sin(u)-\quart[\frac{\dd~u}{\dd X}+\frac{\dd~u}{\dd T}])v_2
,\nono
v_{2,X}&=&i(\frac{\xi}{2}-\frac{cos(u)}{8~\xi}) v_2
+(\frac{i}{8~\xi}sin(u)+\quart[\frac{\dd~u}{\dd X}+\frac{\dd~u}{\dd T}])v_1
,\nono
v_{1,T}&=&-i(\frac{\xi}{2}+\frac{cos(u)}{8~\xi}) v_1
-(\frac{i}{8~\xi}sin(u)+\quart[\frac{\dd~u}{\dd X}+\frac{\dd~u}{\dd T}])v_2
,\nono
v_{2,T}&=&i(\frac{\xi}{2}+\frac{cos(u)}{8~\xi}) v_2
-(\frac{i}{8~\xi}sin(u)-\quart[\frac{\dd~u}{\dd X}+\frac{\dd~u}{\dd T}])v_1
\eea

\noindent where $\ds \xi$ is a complex spectral parameter.

\noindent In eqs.(\ref{akns}),the {\sl X,T} indices denote partial 
derivatives with respect to the physical coordinates.
Consistency between the set of the first two equations and the second two 
equations in eq.(\ref{akns}) with a time independent $\ds \xi$
yields the Sine-Gordon equation for {\sl u(x,t)}.
The solitons of the Sine-Gordon equation correspond to the
bound state sector of the spectrum of the AKNS equations\ci{ablowitz}.
These bound states appear as poles in the transmission coefficients.

Two sets of independent solutions of the AKNS\ci{akns,ablowitz}
equations are defined. The first set $\phi=\ds \left( \begin{array}{c}
v_1\\v_2\\  \end{array}  \right)~~
\bar{\phi}= \left( \begin{array}{c}
\bar{v}_1\\ \bar{v}_2\\ \end{array} \right) $ 

\noindent with boundary conditions at $\ds X\ra -\infty$ 

\bea\label{b1}
\phi&\ra&  \left( \begin{array}{c}
1\\0\\ \end{array} \right) e^{-i \eta(X,T)},\nono
\bar{\phi}&\ra&  \left( \begin{array}{c}
0\\-1\\ \end{array} \right) e^{i \eta(X,T)},\nono
\eta(X,T)&=&(\frac{\xi}{2}-\frac{1}{8\xi})X+(\frac{\xi}{2}+\frac{1}{8\xi})T.
\eea

\noindent The second set $\psi=  \left( \begin{array}{c}
v_1\\v_2\\ \end{array} \right) ~~
\bar{\psi}=  \left( \begin{array}{c}
\bar{v}_1\\ \bar{v}_2\\ \end{array} \right) $ 

\noindent with boundary conditions at $\ds X\ra \infty$ 

\bea\label{b2}
\psi\ra  \left( \begin{array}{c}
0\\1\\ \end{array}  \right) e^{i\eta(X,T)} ,\nono
\bar{\psi}\ra  \left( \begin{array}{c}
1\\0\\ \end{array}  \right)e^{-i\eta(X,T)} .
\eea

\noindent The two sets are connected by the scattering amplitudes 
$\ds a,\bar{a},b,\bar{b}$

\bea\label{connect}
\phi&=&a(\xi)\bar{\psi}+b(\xi)\psi,\nono
\bar{\phi}&=&-\bar{a}(\xi)\psi+\bar{b}(\xi)\bar{\psi}.
\eea

Poles in the transmission amplitude $\ds a(\xi)$ with $\ds Im(\xi)>0$
generate the solitary waves.

The problem of an initial square pulse can be solved analytically. 
The analytical solution can then be compared to the solution
of eqs.(\ref{akns}) for more realistic pulses.

\vspace{2 cm}
Consider a static square pulse

\bea\label{pulse}
u(X,T)&=&A~\Theta(X)~\Theta(w-X),\nono
\frac{\dd u(x,T)}{\dd T}\bigg{|}_{T=0}&=&0,\nono
u_X&=&A~(\delta(X)-~\delta(w-X)),
\eea

\noindent where $\ds \Theta$ denotes the Heaviside step function, $\ds w$
is the width of the pulse and {\sl A} its height, and $\delta$ represents
the Dirac $\delta$ function.
The initial pulse is initially at its maximal (minimal) value.

For $\ds \phi$, the solutions of eq.(\ref{akns}) are

\bea\label{phi}
v_1&=& \Theta (-x)~e^{-i\eta(X,0)}+\Theta (x) \Theta (w-x)
(B~e^{i~\beta~X}+C~e^{-i~\beta~X})+\Theta (x-w)~F~e^{-i\eta(X,0)},\nono
v_2&=& \Theta (x) \Theta (w-x)
(D~e^{i~\beta~X}+E~e^{-i~\beta~X})+\Theta (x-w)~G~e^{i\eta(X,0)},
\eea

\noindent where {\sl B,C,D,E,F,G} are $\ds \xi$ dependent constant
complex amplitudes and $\ds \eta(X,0)$ is defined in eq.(\ref{b1}).
These amplitudes are determined by inserting eq.(\ref{phi}) in eq.(\ref{akns})
 and using $\ds \Theta (0)=\half$

\bea\label{amp}
B&=&\frac{2\delta\eps-(1-\delta^2)(\mu-\beta)}{2\beta~(1+\delta^2)},\nono
C&=&\frac{-2\delta\eps+(1-\delta^2)(\mu+\beta)}{2\beta~(1+\delta^2)},\nono
D&=&\frac{\beta+\mu}{\eps}~B,\nono
E&=&\frac{-\beta+\mu}{\eps}~C,\nono
F&=&\frac{P}{1+\delta^2}~e^{i\eta(w,0)},\nono
P&=&e^{i\beta w}~B~(1-\delta^2+2\delta\frac{\beta+\mu}{\eps})+
e^{-i\beta w}~C~(1-\delta^2+2\delta\frac{\mu-\beta}{\eps}),\nono
G&=&
\bigg(e^{i\beta w}~D~[1-\delta^2-2\delta\frac{\eps}{\beta+\mu}]+
e^{-i\beta w}~E~[1-\delta^2-2\delta\frac{\eps}{-\beta+\mu}]\bigg)
e^{i\eta(w,0)}/(1+\delta^2),\nono
\mu&=&\frac{\xi}{2}-\frac{cos(A)}{8\xi},\nono
\eps &=&\frac{sin(A)}{8\xi},\nono
\beta&=&\sqrt{\mu^2+\eps^2},\nono
\gamma &=& \frac{A}{8}.
\eea

Poles in the transmission amplitude 
correspond to zeroes of the amplitude 
\emph{\sl a} in eq.(\ref{connect}), and to the 
vanishing of the amplitude {\sl P}.
The location of the poles is consequently determined by the implicit equation

\bea\label{pole}
P=e^{i\beta w}~B~(1-\delta^2+2\delta\frac{\beta+\mu}{\eps})+
e^{-i\beta w}~C~(1-\delta^2+2\delta\frac{-\beta+\mu}{\eps})=0.
\eea

\noindent Single breather poles appear in pairs at $\xi=\pm\alpha+i\beta$, with
$\ds \beta\ge 0$
to insure the convergence of the Neumann series of the Volterra integrals  
of the IST.\ci{ablowitz},
For a single breather at rest the spectral parameter obeys
\bea\label{constr}
|\xi |^2=\alpha^2+\beta^2=\quart.
\eea

An initially static pulse can decay to two breathers also.
Momentum conservation requires the breathers to be emitted back to back with
opposite velocity.
A new parameter enters the problem, the velocity of the breathers.
For a breather in motion with velocity {\sl v} the constraint 
of eq.(\ref{constr}) changes. It is easier to find the modified constraint
 in the light cone formulation.
A Lorentz boost of a breather (a scalar field) amounts to
 a Lorentz transformation of the space-time arguments. In light-cone
coordinates with unit speed of light, 
a boost with velocity {\sl v} in the positive {\sl X} direction
transforms the light cone coordinates 

\bea\label{boost}
x\ra\sqrt{\frac{1-v}{1+v}}x,\nono
t\ra\sqrt{\frac{1+v}{1-v}}t.\nono
\eea

\noindent Inspection of the spatial set of the Zakharov-Shabat 
equations in light-cone coordinates for the Sine-Gordon case\ci{akns}

\bea\label{zakh1}
v_{1,x}&=&-i\xi v_1-\half~u_x~v_2,\nono
v_{2,x}&=&i\xi v_2+\half~u_x~v_1,
\eea

\noindent shows that the spectral parameter for a moving breather
transforms as $\ds \xi\ra\sqrt{\frac{1+v}{1-v}}\xi$.
The constraint of eq.(\ref{constr}) now reads $\ds \alpha^2+\beta^2=
\frac{1-v}{4~(1+v)}$. As $\ds v \ra 1$,
one pair of poles moves towards the real $\ds \xi$ axis while the other
pair is pushed up to infinity. 
We here focus on the single breather production case and the double
breather production is dealt with numerically by integrating the Sine-Gordon
equation(see section 3).The analytical treatment of the double breather case 
by means of the Hirota method\ci{hirota} will be addressed in a future work.
For the three breather production from a stationary pulse, one of
the breathers remains at the location of the initial pulse. Therefore, 
three breather production can be treated as the single breather case.

Eq.(\ref{pole}) was derived for the sharp edge square pulse of eq.(\ref{pulse}).
This pulse has infinite energy as may be seen by inserting eq.(\ref{pole}) in 
eq.(\ref{en}), 
due to the $\ds \delta$ function discontinuities at $\ds X=0,\pm \half~w$. 
A more realistic profile 
that still brings about the features of the square profile consists
in the addition of lateral wings, continuous
with the square pulse and its derivative at its edges.
The modified initial pulse (centered at the origin for convenience) reads

\bea\label{pulse1}
u(X,T=0)&=&A~(e^{-z_1^2}~\Theta(-\half~w-X)
+\Theta (X+\half~w)~\Theta (\half~w-X)+e^{-z_2^2}\Theta (X-\half~w)),\nono
z_1&=&\frac{X+\half~w}{\delta w},\nono
z_2&=&\frac{X-\half~w}{\delta w},
\eea

\noindent with $\ds \delta w$, a width parameter for the wings.

\noindent For eq.(\ref{pulse1}) we could not provide analytical solutions.
Eqs.(\ref{akns}) were therefore solved numerically in
order to verify the predictions of eq.(\ref{pole}).
Eqs.(\ref{akns}) are firstly simplified by the substitution

\bea\label{subst}
v_1&\ra&v_1~e^{-i \rho(X)},\nono
v_2&\ra&v_2~e^{i \rho(X)},\nono
\rho(X)&=&\half(\xi~X-\frac{\int{cos(u(X))~dX}}{8~\xi}).
\eea

\noindent The spatial set of the AKNS equations(\ref{akns}) now becomes
\bea\label{akns1}
v_{1,X}&=&(\frac{i}{8~\xi}sin(u)-
\quart[\frac{\dd~u}{\dd X}+\frac{\dd~u}{\dd T}])~v_2~e^{2~i\rho(X)},\nono
v_{2,X}&=&
(\frac{i}{8~\xi}sin(u)
+\quart[\frac{\dd~u}{\dd X}+\frac{\dd~u}{\dd T}])~v_1~e^{-2~i\rho(X)},
\eea
with boundary conditions at $\ds X \ra-\infty$, $\ds v_1\ra~1,~v_2\ra~0$.

We solved eqs.(\ref{akns1}) using a fourth order Runge-Kutta method with 
double precision in order to achieve a maximal error of 1\%. 
The spatial grid was fixed at $\ds \delta X=.0002$ with $\ds -10<X<10$.
Poles in the transmission matrix were located by searching for
$\ds D=abs(v_1(\ra \infty))<10^{-5}$.
For the wings width parameter we took $\delta w=0.05$, a small enough
value to compare to the analytical results of the square pulse
above.

Figure 1 shows area plots for breathers production. The single breather 
regions were found using the analytical formula of eq.(\ref{pole}), curve
labeled AKNS analytical, 
solutions of eq.(\ref{akns1}), curve labeled AKNS numerical, and integration
of the Sine-Gordon equation, curve labeled Sine-Gordon numerical.
The thresholds for two breathers were obtained from the
Sine-Gordon equation integration, 
while three breather thresholds followed from both
the AKNS numerical treatment of eq.(\ref{akns1}) and the square pulse 
formula of eq.(\ref{pole}).
The curve labeled {\sl Sine-Gordon integration}
does not match the AKNS prediction. This is due to the fact that
the observability of the breathers demands a minimal amplitude for
them such that they do not blend into the phonon continuum. The thresholds
found by the AKNS equations correspond to $\xi_{thresh}=0.5+i~0$, 
hence a vanishing breather height. The Sine-Gordon integration
curve was drawn for breather amplitudes corresponding to a pole at $\xi=0.499+
i~0.0316$ at which the breather becomes clearly visible.
We searched for poles in the parameter region $\ds 0<A<10 ,0.5<w<10$, so that
$\ds \delta w$ stayed reasonably smaller than the width of the 
pulse. 

\begin{figure}
\epsffile{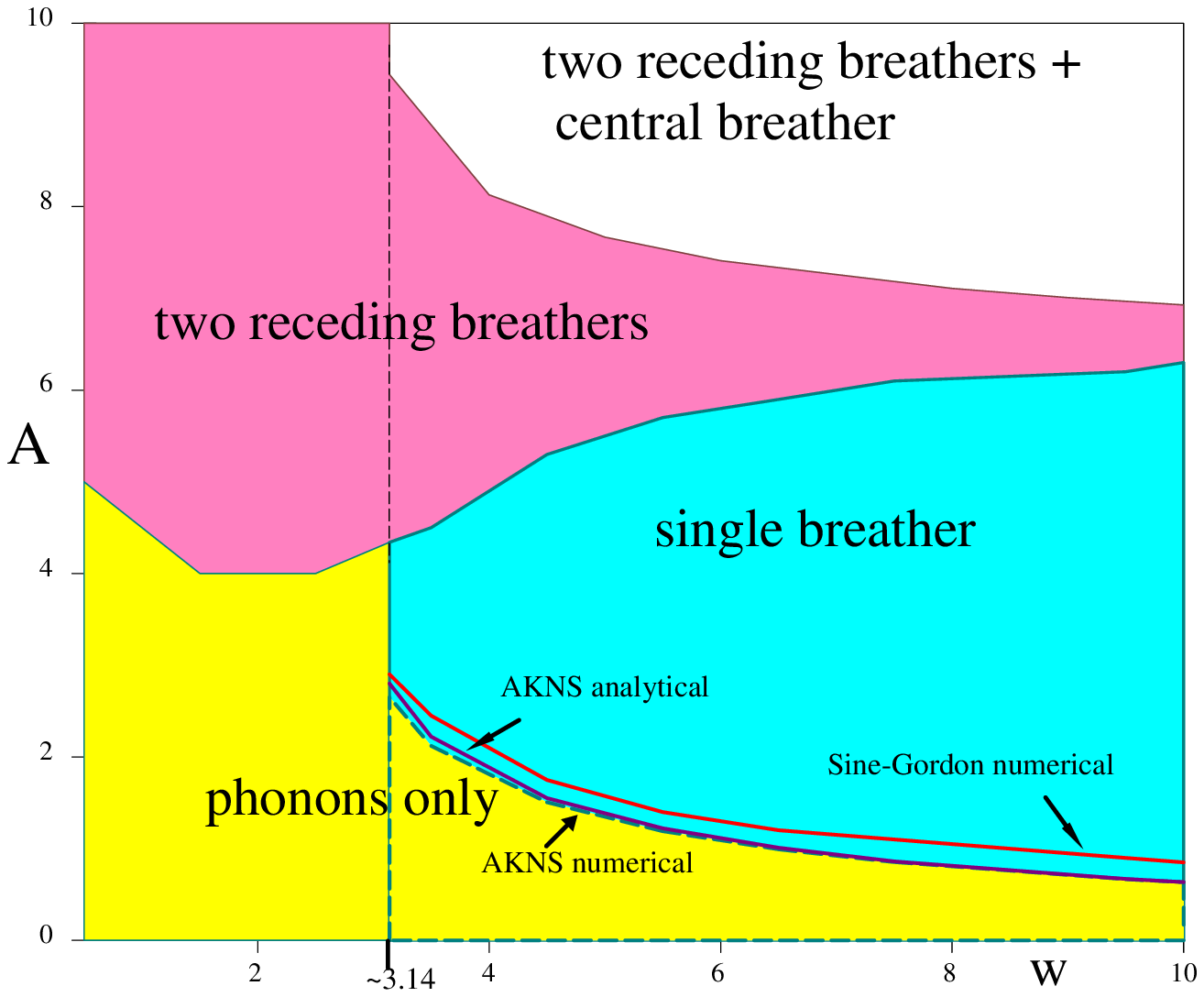}
\caption{ Area plot for the parameter space leading to the production of 
Sine-Gordon breathers}
\label{fig1}
\end{figure}

The agreement between the square pulse predictions and the full AKNS
treatment is satisfactory. From the plot we see that there is no single
or triple breather production unless $\ds w>\approx\pi$.
Figure 2 depicts a typical solution of eqs.(\ref{akns1}) at a value of the
spectral parameter corresponding to a pole.
The wave functions are quite constant in each region as in the ansatz of
eq.(\ref{phi}) corresponding to the square pulse.

\begin{figure}
\epsffile{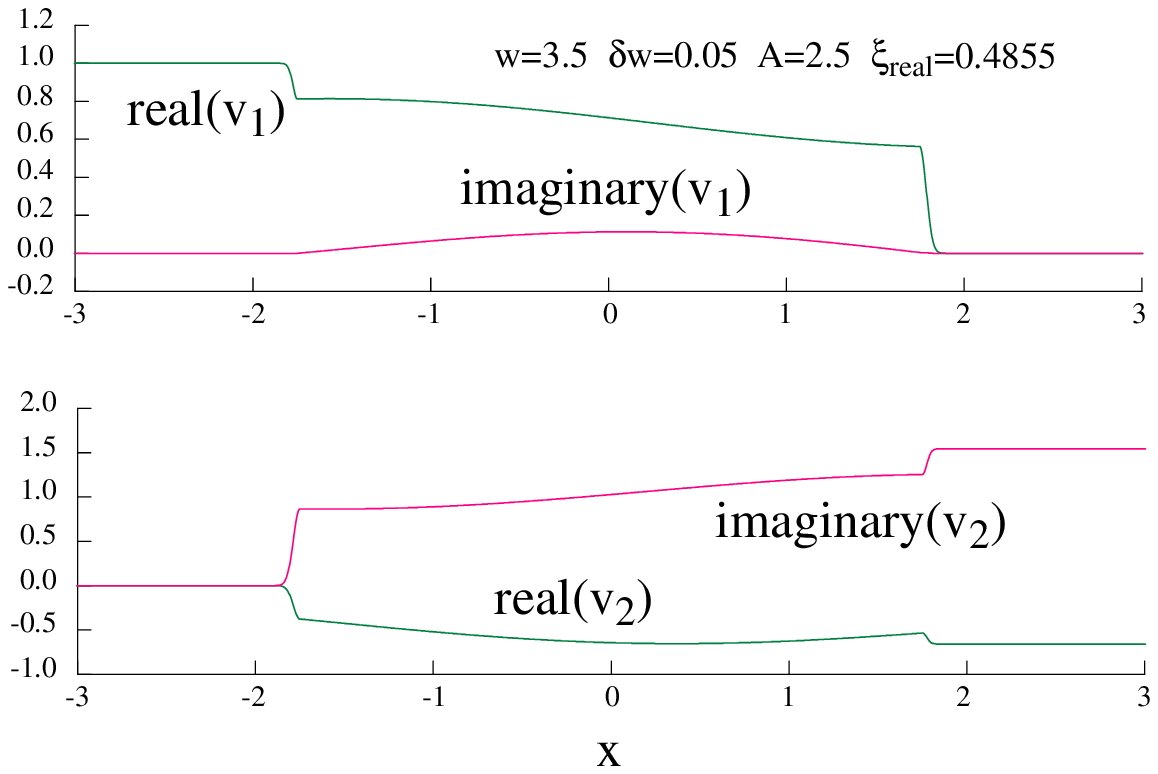}
\caption{$v_1$ and $v_2$ of eq.(\ref{akns1}) as a function of distance 
for a value of $\xi$ correponding to a pole in the transmission matrix.}
\label{fig2}
\end{figure}

In the next section we depict the various breather production thresholds,
and compare the breather profile to the theoretical expression.

\section{\sl Decay of a square pulse, numerical simulations}

\begin{figure}
\epsffile{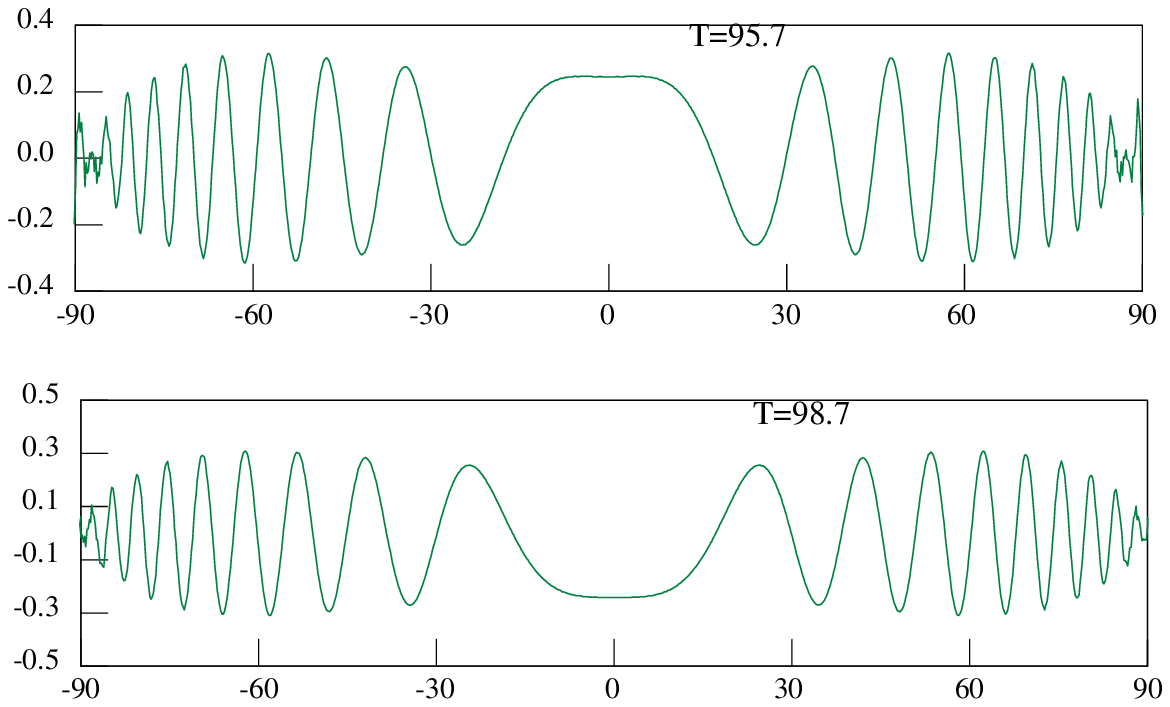}
\caption{ Sine-Gordon solution for an initial pulse with $A=2.5~w=2.5$, below
the width and height thresholds.}
\label{fig3}
\end{figure}
 
The numerical integration of the Sine-Gordon equation, was performed 
 implementing the numerical code described in \ci{k1}.
The spatial and temporal steps needed to achieve 1\% accuracy in the
final energy were $\ds dx=0.02, dt=0.01$. 
\begin{figure}
\epsffile{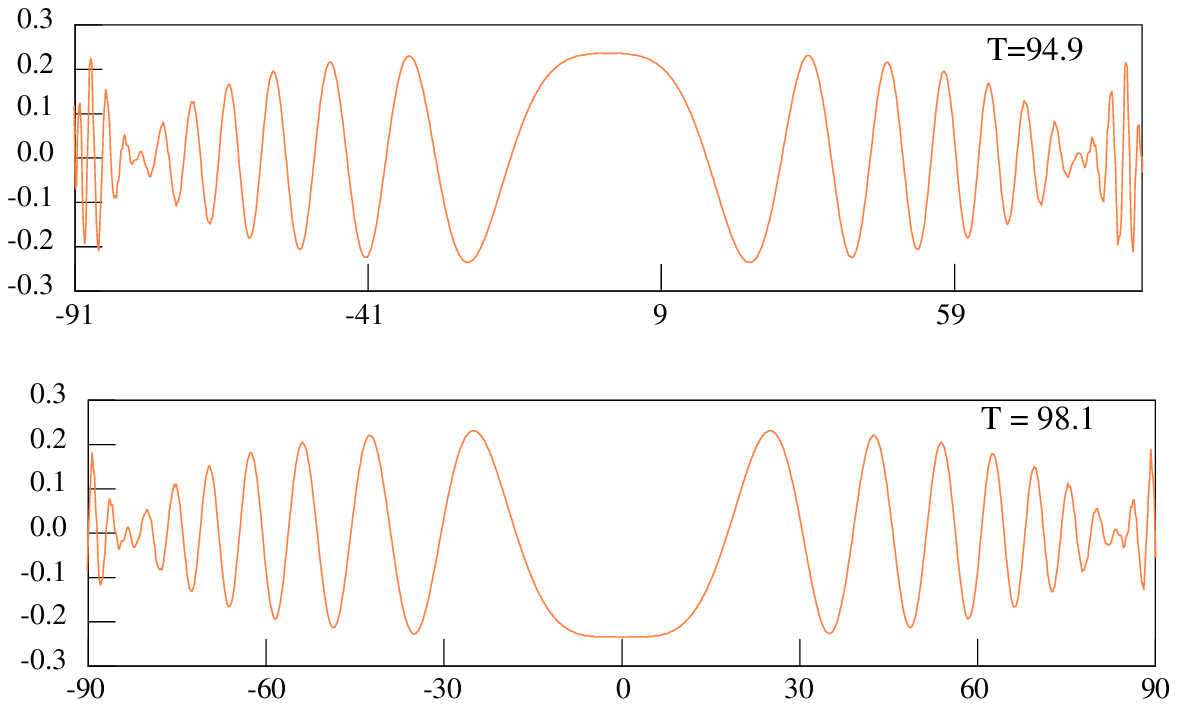}
\caption{Sine-Gordon solution for an initial pulse with $A=1.5~w=3.5$, above the
width threshold and below the height threshold.}
\label{fig4}
\end{figure}
The spatial extent of the integration region was taken to
be $\ds X_{max}=600$, with vanishing fields at infinity.
In the examples depicted below we used $\ds \delta w=0.05$.

In section 2 we found two thresholds
for single breather production from a square pulse.
The width threshold reads $\ds w_{thresh}\approx\pi$. The height threshold
can be obtained from the curve in figure 1.
Figure 3 depicts a typical case below the width and height thresholds.
A trail of phonons emanating from the origin is clearly observed, as well
as the remnant of the original profile.
The angular frequency of the oscillation of the central peak is greater than
the upper bound for the breather pole $\alpha=\frac{\pi}{\bf T}=0.523~>0.5$,
with $\bf T$, the oscillation period. 
Moreover, the central peak does not have the shape of a breather.
Figure 4 depicts a case below the height threshold 
and above the width threshold.
Figure 5 shows a case above both width and height thresholds.
\begin{figure}
\epsffile{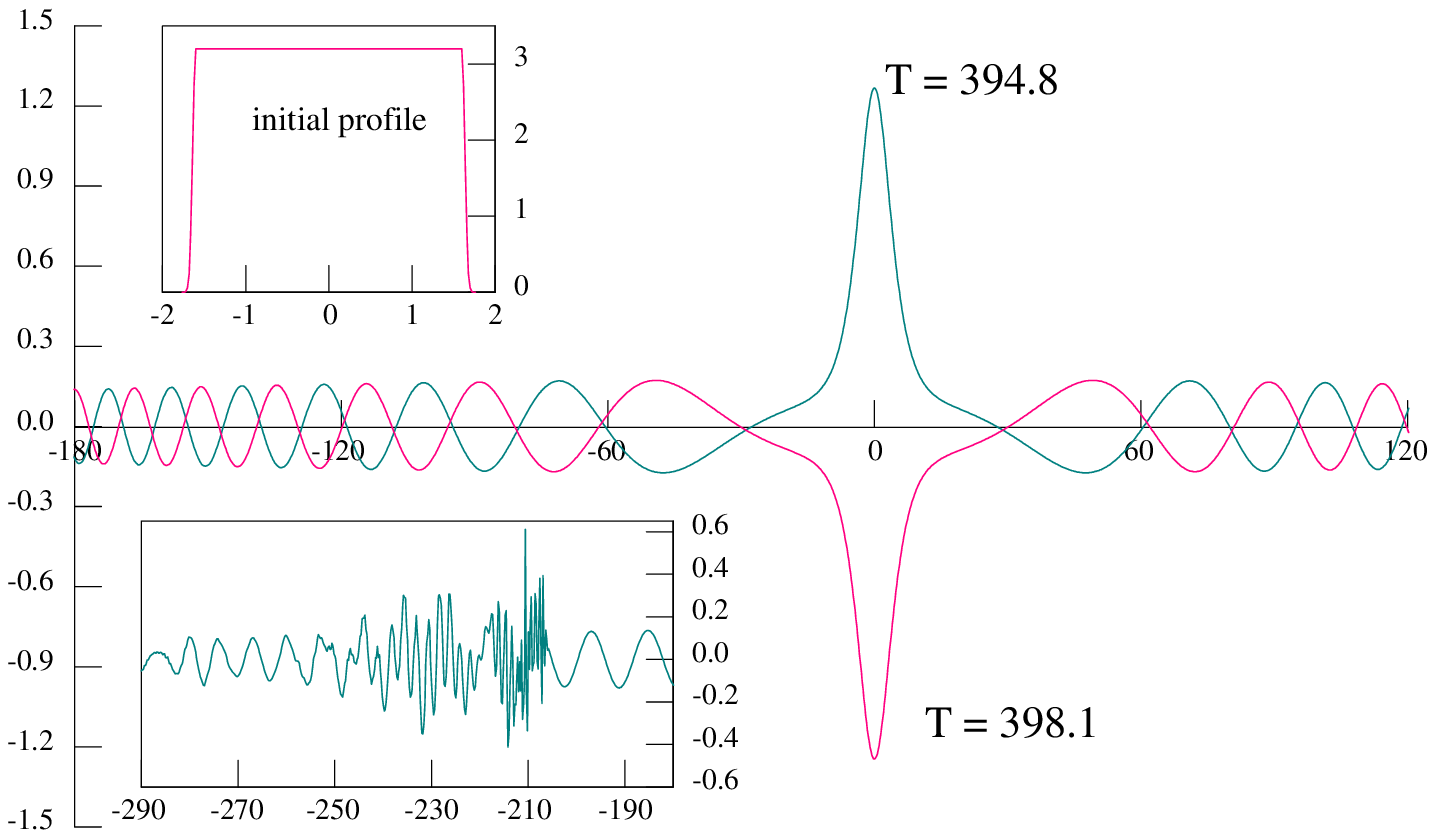}
\caption{Sine-Gordon solution for an initial pulse with $A=3.2~w=3.2$, above the
width and height thresholds.}
\label{fig5}
\end{figure}
In figures 3 and 4 only phonons are seen to 
emanate from the central region, while figure 5 shows
a prominent peak centered at the origin and a small trail of phonons 
in both directions.
From the graph of figure 5 the oscillation period
is found to be {\bf T}=6.6. Using $\ds \xi=\alpha+i\beta$, with
$\ds \beta=\sqrt{0.25-\beta^2}$, and $\ds \alpha=\frac{\pi}{\bf T}$, we can
obtain the location of the pole and evaluate eq.(\ref{pole}) with 
the amplitude and the height of the pulse as in figure 5.
With these data, eq.(\ref{pole}) yields $ P =2~10^{-3}$, a value
close to zero, as compared to $P\approx 1$ for $\xi$ far from a pole location.

\noindent 
\begin{figure}
\epsffile{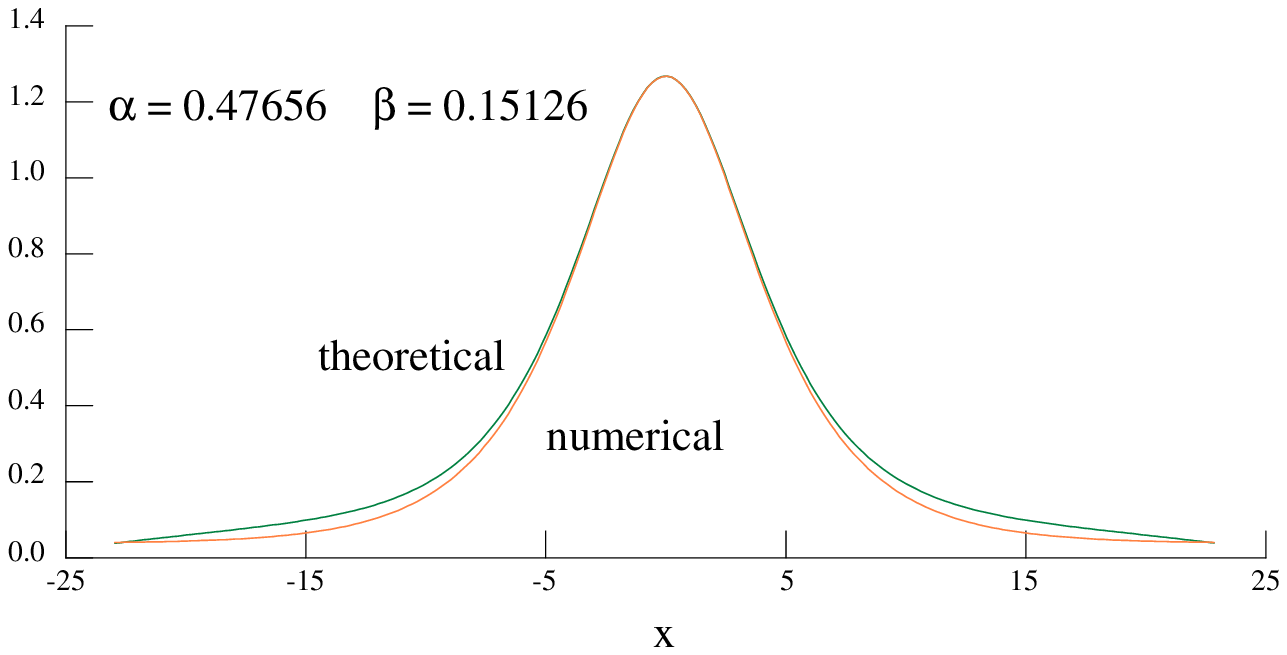}
\caption{Central peak of the Sine-Gordon solution of figure 5
compared to eq.(\ref{breather}).}
\label{fig6}
\end{figure}

We can now compare the shape of the 
central peak in figure 5, presumed to be a breather,
with the theoretical expression 
\begin{figure}
\epsffile{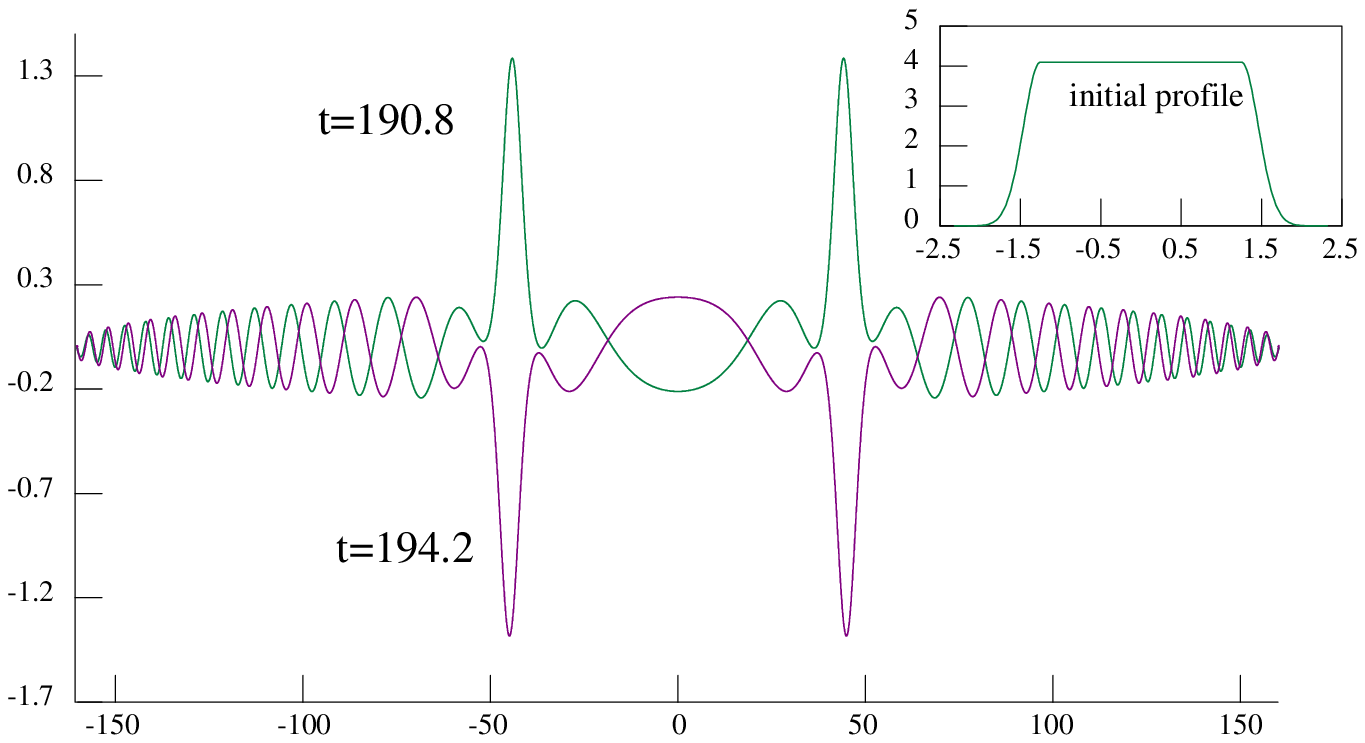}
\caption{ Sine-Gordon equation solution for an 
initial pulse with $A=4.1,~w=2.5$, below the single breather width threshold.}
\label{fig7}
\end{figure}

\bea\label{breather}
u(X,T)&=&4~tan^{-1}\bigg(\frac{\frac{\beta}{\alpha}~U}{V}\bigg),\nono
U&=&cos(2\alpha~(T+T_0)),\nono
V&=&cosh(2\beta~X).
\eea

\noindent This correspondence is depicted in figure 6. Except for small phonon
contributions, the profiles fit nicely. The central peak is a breather.

\begin{figure}
\epsffile{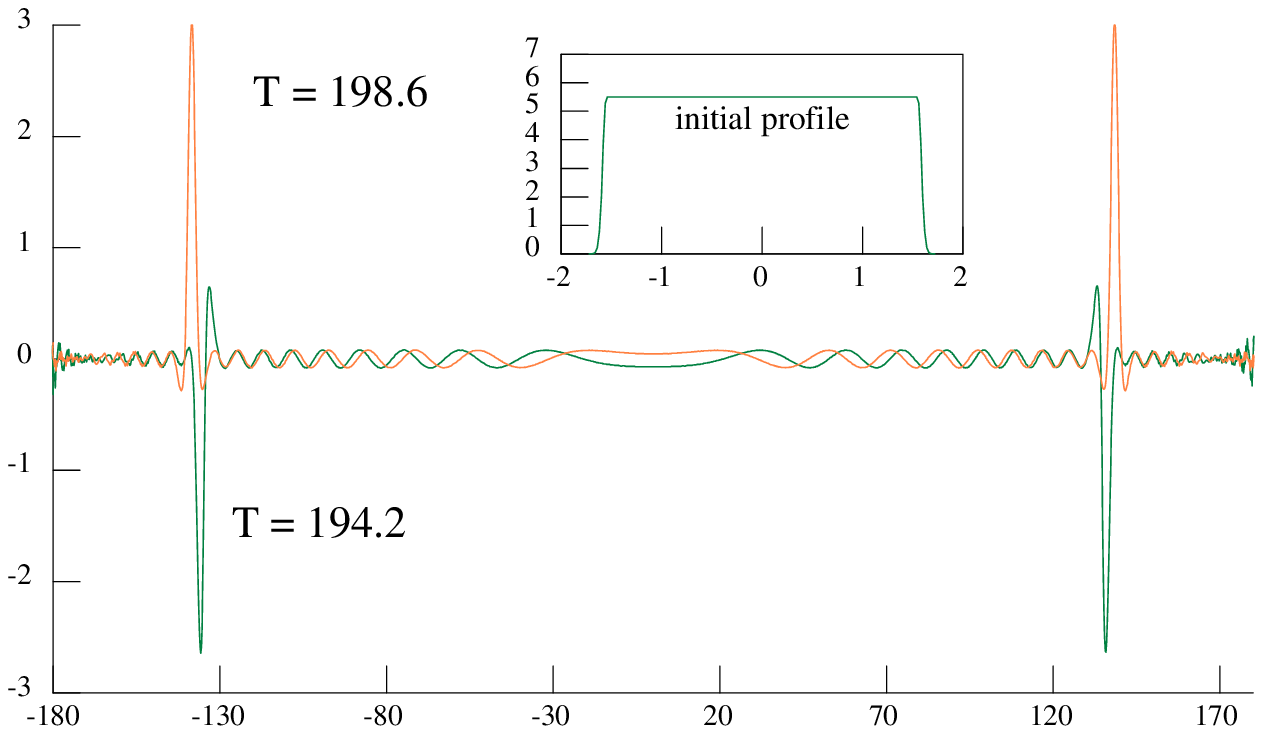}
\caption {Sine-Gordon equation solution for an 
initial pulse with $\ds A=5.5,~w=3.1$, above the single breather 
width threshold}
\label{fig8}
\end{figure}

Figures 7 and 8 show numerical simulations for the two breather region
of figure 1. A pair of back to back 
breathers receding from the center is observed.

\begin{figure}
\epsffile{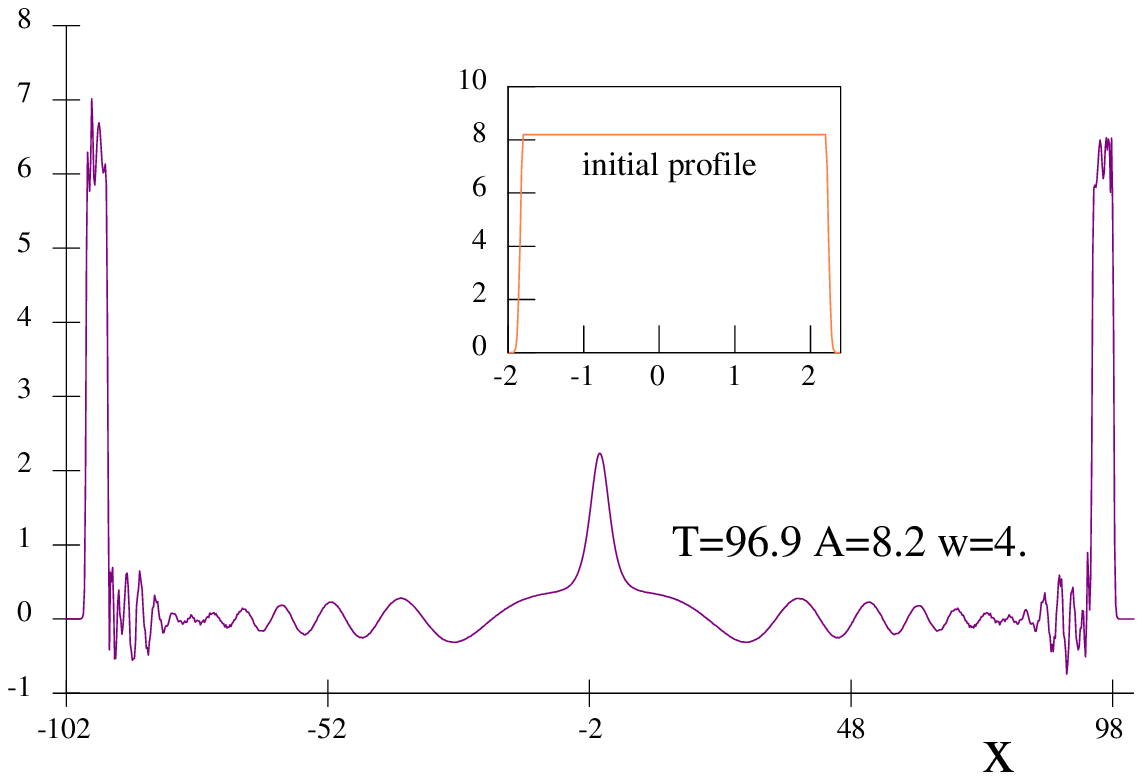}
\caption {Sine-Gordon equation solution for an 
initial pulse with $\ds A=8.2,~w=4$}
\label{fig9}
\end{figure}

Figure 9 shows a three breather situation.

\section{Conclusions}
In the present investigation we considered the decay of
an initially nonvanishing static square pulse. The results
complement those of Kivshar et al.\ci{kiv1,kiv2}.
However, contrarily to Kivshar et al.\ci{kiv2} 
we could not find any case for which
a static square pulse leads to soliton-antisoliton pair production
 regardless of its width, height or energy. A different
kind of initial pulse or a time dependent profile is apparently needed.

Among the rich structure of the decaying pulse,  the phonon 
spectrum certainly deserves attention. We found that in general
the main features of the spectrum can be reproduced
with the long time analytical expressions of Segur and Ablowitz\ci{segur1},
however, the fine details do not match existing expressions in the
literature.

We only touched upon the double breather problem and will deal with
it in the future.

We conclude with a remark concerning the experimental relevance of the present 
work. 
The initial pulse configuration we chose may be realized experimentally in long
Josephson junctions of regular or high $T_c$ superconductors by
injecting two oppositely directed  bias currents
at a distance {\sl w} corresponding to the pulse width.
By regulating the current
intensity and the distance between injection points a well
defined breather with a fixed oscillation period may be produced, with
possible technological implications.

\end{document}